\documentclass[11pt,twoside,a4paper,cmspaper,final,collab]{cms-tdr}
\def\svnVersion{9283d2b}\def\svnDate{2026/08/19}\def\cmsCernNoTag{CERN-EP-2026-189}\def\cmsCernDate{\today}\def\cmsMessage{Submitted to Physical Review Letters}
\begin{document}\cmsNoteHeader{TOP-25-015}

\newcommand*{\tmp}[4]{%
  #4^{#1#3}_{#2}%
}

\newcommand*{\eftOp}[4]{\ensuremath{%
    {#4    
    \ifx\empty#2\empty\else_{#2}\fi^{#1#3}}%
}}

\newcommand*{\qq}[4][]{\tmp{#2}{#3}{#4}{#1{O}}}

\newcommand{\FDF}{(\varphi^\dagger i\!\!\overleftrightarrow{D}_\mu\varphi)}
\newcommand{\hc}[1]{{}^\ddagger #1}

\newcommand{\Plepton}{\Pell}
\newcommand{\PAlepton}{\PAell}
\newcommand{\ttH}{\ensuremath{\ttbar\PH}\xspace}
\newcommand{\ttZ}{\ensuremath{\ttbar\PZ}\xspace}
\newcommand{\ttW}{\ensuremath{\ttbar\PW}\xspace}
\newcommand{\tZq}{\ensuremath{\PQt \PZ q}\xspace}
\newcommand{\tWZ}{\ensuremath{\PQt \PW \PZ}\xspace}
\newcommand{\ttG}{\ensuremath{\ttbar\PGg}\xspace}
\newcommand{\sigmaii}{\ensuremath{\sigma_{i\eta i\eta }}\xspace}
\newcommand{\WG}{\ensuremath{\PW\PGg\text{+jets}}\xspace}
\newcommand{\tqG}{\ensuremath{\PQt\Pq\PGg}\xspace}
\newcommand{\ZG}{\ensuremath{\PZ\PGg\text{+jets}}\xspace}

\newcommand{\sW}{\ensuremath{s_W}}
\newcommand{\cW}{\ensuremath{c_W}}

\newcommand{\ctpij}  {\eftOp{}{\PQt\varphi}{(ij)}{c}\xspace}
\newcommand{\ctpji}  {\eftOp{}{\PQt\varphi}{(ji)}{c}\xspace}
\newcommand{\cpQMij} {\eftOp{-}{\varphi Q}{(ij)}{c}\xspace}
\newcommand{\cptij}  {\eftOp{}{\varphi\PQt}{(ij)}{c}\xspace}
\newcommand{\ctZij}  {\eftOp{}{\PQt\PZ}{(ij)}{c}\xspace}
\newcommand{\ctAij}  {\eftOp{}{\PQt \ensuremath{\rm A}}{(ij)}{c}\xspace}
\newcommand{\ctGij}  {\eftOp{}{\PQt \cPG}{(ij)}{c}\xspace}

\newcommand{\cQlMij} {\eftOp{-}{Q\Plepton}{(\Plepton)(ijkl)}{c}\xspace}
\newcommand{\cQeij}  {\eftOp{}{Q\Pe}{(\Plepton)(ijkl)}{c}\xspace}
\newcommand{\ctlij}  {\eftOp{}{\PQt \Plepton}{(\Plepton)(ijkl)}{c}\xspace}
\newcommand{\cteij}  {\eftOp{}{\PQt \Pe}{(\Plepton)(ijkl)}{c}\xspace}
\newcommand{\ctlSij} {\eftOp{S}{\PQt}{(\Plepton)(ijkl)}{c}\xspace}
\newcommand{\ctlTij} {\eftOp{T}{\PQt}{(\Plepton)(ijkl)}{c}\xspace}

\newcommand{\ctp}  {\eftOp{}{\PQt\varphi}{}{c}\xspace}
\newcommand{\cpQM} {\eftOp{-}{\varphi Q}{}{c}\xspace}
\newcommand{\cpt}  {\eftOp{}{\varphi\PQt}{}{c}\xspace}
\newcommand{\ctZ}  {\eftOp{}{\PQt\PZ}{}{c}\xspace}
\newcommand{\ctA}  {\eftOp{}{\PQt \ensuremath{\rm A}}{}{c}\xspace}
\newcommand{\ctG}  {\eftOp{}{\PQt \cPG}{}{c}\xspace}

\newcommand{\cQlM} {\eftOp{-}{Q\Plepton}{(\Plepton)}{c}\xspace}
\newcommand{\cQe}  {\eftOp{}{Q\Pe}{(\Plepton)}{c}\xspace}
\newcommand{\ctl}  {\eftOp{}{\PQt \Plepton}{(\Plepton)}{c}\xspace}
\newcommand{\cte}  {\eftOp{}{\PQt \Pe}{(\Plepton)}{c}\xspace}
\newcommand{\ctlS} {\eftOp{S}{\PQt}{(\Plepton)}{c}\xspace}
\newcommand{\ctlT} {\eftOp{T}{\PQt}{(\Plepton)}{c}\xspace}

\newcommand{\twolss}{\ensuremath{2\Pell \textrm{ss}}\xspace}
\newcommand{\threel}{\ensuremath{3\Pell}\xspace}
\newcommand{\threeloffZ}{\ensuremath{3\Pell\text{ off-}\PZ}\xspace}
\newcommand{\threelonZ}{\ensuremath{3\Pell\text{ on-}\PZ}\xspace}

\cmsNoteHeader{TOP-25-015}

\title{Search for top quark flavor-changing neutral currents in multilepton final states using an effective field theory approach in proton--proton collisions at \texorpdfstring{$\sqrt{s}=13\TeV$}{sqrt(s) = 13 TeV}}

\author[cern]{A. Cern Person}
\date{\today}

\abstract{
We present a search for flavor-changing neutral current (FCNC) interactions involving the top quark in production and decay. This analysis establishes the first global search in the top quark FCNC sector within the standard model effective field theory where all relevant operators are constrained simultaneously in a single fit to data, providing a framework for future measurements and combinations at the LHC. The analysis uses proton--proton collision data collected at $\sqrt{s}=13\TeV$ by the CMS detector, corresponding to an integrated luminosity of 138\fbinv. Signal contributions are probed in final states with two same-sign charged leptons or three leptons. Multivariate discriminants are used to separate signal from background, and their distributions enter a likelihood fit to determine 95\% confidence intervals for 24 Wilson coefficients. No evidence for FCNC interactions is observed. These results provide the first simultaneous constraints on the complete operator set and place the most stringent limits to date on FCNC top quark decay branching fractions via two-quark--two-lepton interactions.
}

\hypersetup{%
pdfauthor={CMS Collaboration},%
pdftitle={Search for top quark flavor-changing neutral currents (FCNC) in multilepton final states using an effective field theory approach in proton-proton collisions at sqrt(s) = 13 TeV},%
pdfsubject={CMS},%
pdfkeywords={CMS, TOP, FCNC, EFT}}

\maketitle 

In the standard model (SM), flavor-changing neutral current (FCNC) processes are highly suppressed by the Glashow--Iliopoulos--Maiani mechanism~\cite{Glashow:1970gm}, with branching fractions predicted to be at the level of $\mathcal{O}(10^{-12})$ or smaller~\cite{AguilarSaavedra:2004wm}.
Several extensions of the SM introduce additional sources of FCNCs, particularly in the top (\PQt) quark sector. Examples include $R$-parity violating supersymmetric models~\cite{Yang:1997dk,Delepine:2004hr,Cao:2007dk,Yang:2018utw} and models with extended Higgs sectors, such as two-Higgs-doublet models~\cite{Atwood:1996vj,Bejar:2000ub}.
These scenarios can enhance the rates of top quark FCNC processes to values potentially observable at the CERN LHC.
The observation of rare FCNC processes would therefore provide unambiguous evidence for physics beyond the SM (BSM)~\cite{AguilarSaavedra:2004wm}.

The ATLAS and CMS Collaborations have performed multiple searches in various final states to probe top quark FCNC interactions mediated by the Higgs boson~\cite{ATLAS:2024mih,CMS:2024ubt}, gluon~\cite{ATLAS:2021amo,CMS:2016uzc}, photon~\cite{ATLAS:2022per,CMS:2023bjm}, or $\PZ$ boson~\cite{ATLAS:2023qzr,CMS:2017wcz}. 
These studies typically consider FCNC processes mediated by a single particle at a time, neglecting potential correlations or interference effects. In addition, they do not account for the possibility of FCNCs arising directly from heavy BSM states.

This letter presents the first search that simultaneously includes all sensitive top quark FCNC interactions within the framework of the SM effective field theory (SMEFT), using a data set of proton--proton ($\Pp\Pp$) collisions at $\sqrt{s} = 13\TeV$ collected by the CMS experiment in 2016--2018 at the LHC. The analysis targets final states containing either two  leptons ($\Pell = \Pe, \PGm$) with the same electric charge (same-sign) or three leptons, collectively referred to as multilepton final states.

Within the SMEFT framework, deviations from the SM are described by extending the SM Lagrangian with higher-dimensional, gauge-invariant operators constructed from SM fields:
\begin{equation}
\mathcal{L}_{\mathrm{EFT}} = \mathcal{L}_{\mathrm{SM}} + \sum_{d,i} \frac{c^{d}_{i}}{\Lambda^{d-4}}\, \mathcal{O}^{d}_{i},
\end{equation}
where $\mathcal{O}^{d}_{i}$ are dimension-$d$ operators, $c^{d}_{i}$ are dimensionless Wilson coefficients (WCs), and $\Lambda$ is the scale of BSM physics. The leading contributions that conserve baryon and lepton number arise from dimension-six operators~\cite{AguilarSaavedra:2018nen}.
In contrast to previous SMEFT searches, which typically assume minimal flavor violation~\cite{CMS:2023xyc,ATLAS:2025yww,CMS:2025ugn} and thereby forbid FCNCs, this analysis targets a reduced set of operators that directly mediate FCNC interactions in the top quark sector, as proposed in Ref.~\cite{Durieux:2014xla}.
In multilepton final states, FCNC interactions can arise from operators involving either two quark and bosonic fields or two quark and two lepton fields. This letter presents the first direct probe of the latter class of operators.

The dimension-six operators considered in this analysis, together with their corresponding WCs, are listed in Table~\ref{tab:eftOperators}. In the general SMEFT framework, the WCs associated with left- and right-handed top quark fields are independent parameters. To reduce the number of free parameters in the fit, we assume equal coupling strengths for the two chiral structures. For example, for the coefficient \ctp\ we impose $\ctpij=\ctpji$ and define the fitted parameter as $\ctp^{2}=\ctpij^{2}+\ctpji^{2}$. Similarly, for two-quark--two-lepton operators, a common WC is assumed for the three charged-lepton flavors. Therefore, we probe 24 WCs: 12 associated with the $\PQt\PQu$ FCNC interactions and 12 associated with the $\PQt\PQc$ FCNC interactions.

\begin{table}[hbt!]
\centering
\topcaption{Operators that can generate FCNC vertices at order $1/\Lambda^2$ and are considered in this analysis. 
Details  on the operators, conventions, and symbol definitions are provided in Ref.~\cite{AguilarSaavedra:2018nen}.}  
\begin{scotch}{lll}
Operator                          & Definition                                                                    & Wilson coefficient         \\ \hline 
\multicolumn{3}{c}{Two-quark + boson operators} \\ \hline 
\noalign{\vskip 0.1cm}
$\hc{\qq{}{\PQu\varphi}{(ij)}}$      & $\PAQq_i \PQu_j\tilde\varphi\: (\varphi^{\dagger}\varphi)$                     & $ \ctpij        $   \\ 
$\qq{1}{\varphi \PQq}{(ij)}$           & $\FDF (\PAQq_i\Pgg^\PGm \PQq_j)$                                              & $ \cpQMij         $   \\ 
$\qq{}{\varphi \PQu}{(ij)}$            & $\FDF (\PAQu_i\Pgg^\PGm \PQu_j)$                                              & $ \cptij                 $   \\  
$\hc{\qq{}{\PQu\PW}{(ij)}}$              & $(\PAQq_i\sigma^{\PGm\PGn}\tau^I\PQu_j)\:\tilde{\varphi}\PW_{\PGm\PGn}^I$            & $\cW \ctAij - \sW \ctZij$           \\  
$\hc{\qq{}{\PQu\PB}{(ij)}}$              & $(\PAQq_i\sigma^{\PGm\PGn} \PQu_j)\quad\:\tilde{\varphi}\PB_{\PGm\PGn}$              & $ \cW \ctZij + \sW \ctAij  $\\ 
$\hc{\qq{}{\PQu G}{(ij)}}$              & $(\PAQq_i\sigma^{\PGm\PGn}T^A\PQu_j)\:\tilde{\varphi}G_{\PGm\PGn}^A$               & $ \ctGij        $   \\ 
\noalign{\vskip 0.1cm}
\hline 
\multicolumn{3}{c}{Two-quark-two-lepton operators } \\ 
\hline
\noalign{\vskip 0.1cm}
$\qq{1}{\Plepton\PQq}{(ijkl)}$                &$(\PAlepton_i\Pgg^\PGm \Plepton_j)(\PAQq_k\Pgg^\PGm \PQq_\Plepton)$                             & $ \cQlMij        $   \\ 
$\qq{}{\Plepton\PQu}{(ijkl)}$                 &$(\PAlepton_i\Pgg^\PGm \Plepton_j)(\PAQu_k\Pgg^\PGm \PQu_\Plepton)$                             & $ \ctlij                 $   \\  
$\qq{}{\Pe\PQq}{(ijkl)}$                 &$(\bar \Pe_i\Pgg^\PGm \Pe_j)(\PAQq_k\Pgg^\PGm \PQq_\Plepton)$                             & $ \cQeij                 $   \\ 
$\qq{}{\Pe\PQu}{(ijkl)}$                 &$(\bar \Pe_i\Pgg^\PGm \Pe_j)(\PAQu_k\Pgg^\PGm \PQu_\Plepton)$                             & $ \cteij                 $   \\  
$\hc{\qq{1}{\Plepton\Pe\PQq\PQu}{(ijkl)}}$         &$(\PAlepton_i \Pe_j)\;\varepsilon\;(\PAQq_k \PQu_\Plepton)$                                  & $ \ctlSij   $   \\ 
$\hc{\qq{3}{\Plepton\Pe\PQq\PQu}{(ijkl)}}$         &$(\PAlepton_i \sigma^{\PGm\PGn} \Pe_j)\;\varepsilon\;(\PAQq_k \sigma_{\PGm\PGn} \PQu_\Plepton)$  & $ \ctlTij     $   \\ \noalign{\vskip 0.1cm}
\end{scotch}
\label{tab:eftOperators}
\end{table}

\begin{figure}[hbt!]
  \centering
      \includegraphics[width=0.48\textwidth]{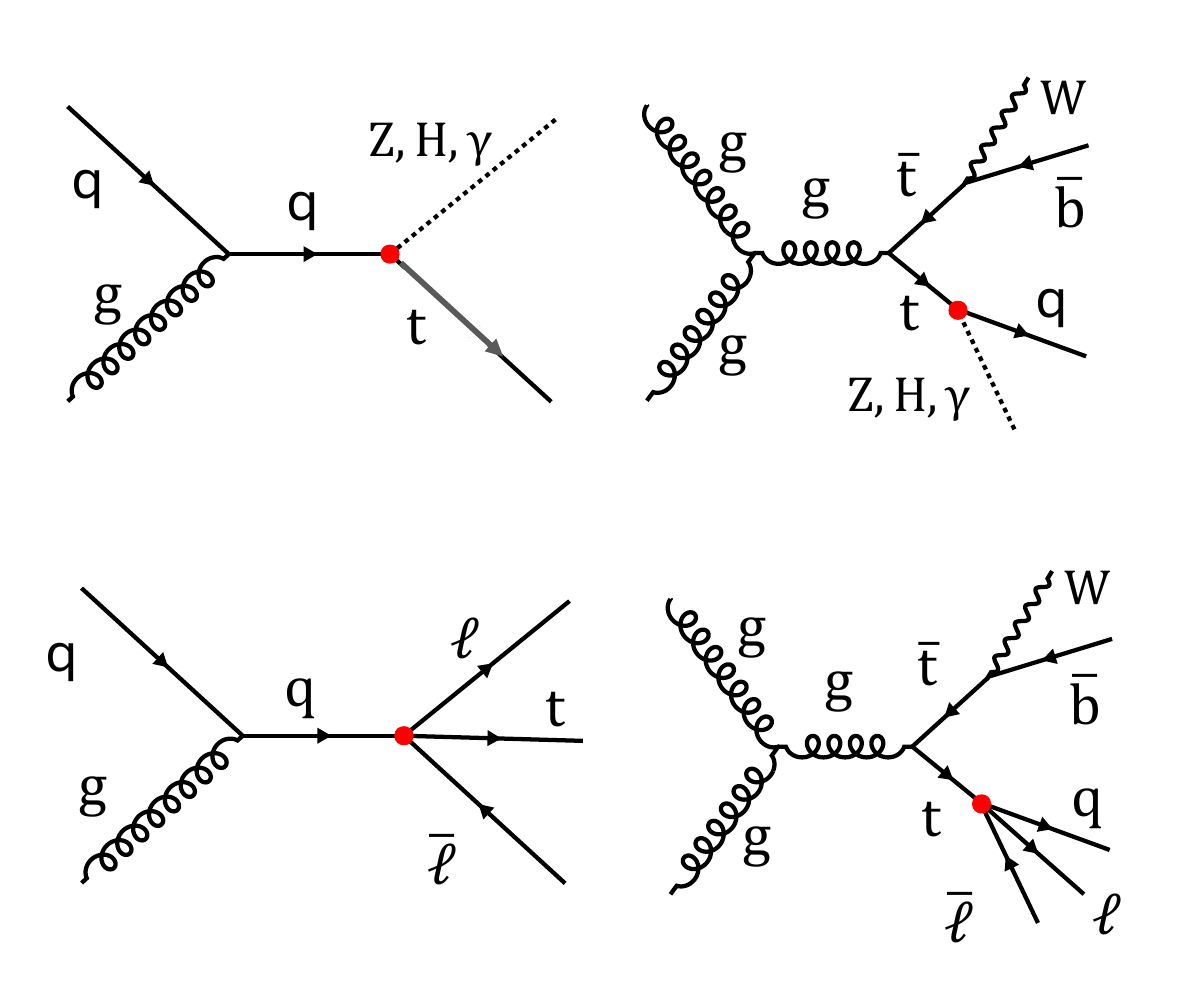}
    \caption{ Representative Feynman diagrams for single top quark production (left) and top quark decays in \ttbar events (right) via FCNC interactions. The two upper panels illustrate the two-quark--plus-boson processes, whereas the two lower panels show the two-quark--two-lepton processes. The FCNC vertex is highlighted with a red circle.
    \label{Feyn}}
\end{figure}

In $\Pp\Pp$ collisions, FCNC interactions of the top quark can appear either in single top quark production (``production mode'') or in top quark decays (``decay mode''). In the multilepton final states considered here, these interactions proceed via a Higgs boson, photon, gluon, \PZ boson, or directly through a lepton pair. Representative Feynman diagrams for both modes are shown in Fig.~\ref{Feyn}.

The WCs $\ctG$ and $\ctA$ parameterize FCNC interactions mediated by a gluon and a photon, respectively. Processes involving  $\ctG$ affect only the production mode; $\ctA$ contributes solely through off-shell photon exchange. Both are more tightly constrained in dedicated searches for single top quark production in association with a jet or a photon~\cite{ATLAS:2021amo,CMS:2016uzc,ATLAS:2022per,CMS:2023bjm}. In this analysis, their contributions are nevertheless included for completeness, despite the substantially lower sensitivity of the multilepton final state relative to single-lepton channels.
 
The CMS apparatus~\cite{CMS:2008xjf,CMS:2023gfb} is a multipurpose, nearly hermetic detector, designed to trigger on~\cite{CMS:2020cmk,CMS:2016ngn,CMS:2024aqx} and identify electrons, muons, photons, and (charged and neutral) hadrons~\cite{CMS:2020uim,CMS:2018rym,CMS:2014pgm}. Its central feature is a superconducting solenoid of 6\unit{m} internal diameter, providing a magnetic field of 3.8\unit{T}. Within the solenoid volume are a silicon pixel and strip tracker, a lead tungstate crystal electromagnetic calorimeter (ECAL), and a brass and scintillator hadron calorimeter, each composed of a barrel and two endcap sections. Forward calorimeters extend the pseudorapidity ($\eta$) coverage provided by the barrel and endcap detectors. Muons are reconstructed using gas-ionization detectors embedded in the steel flux-return yoke outside the solenoid. More detailed descriptions of the CMS detector, together with a definition of the coordinate system used and the relevant kinematic variables, can be found in Refs.~\cite{CMS:2008xjf,CMS:2023gfb}.

A global ``particle-flow'' (PF) algorithm~\cite{CMS:2017yfk} aims to reconstruct all individual particles in an event (PF candidates), combining information provided by the inner tracker and the calorimeters, 
with data from the muon detectors. The missing transverse momentum vector \ptvecmiss is computed as the negative vector sum of the transverse momenta
of all PF candidates in an event, and its magnitude is denoted as \ptmiss~\cite{CMS:2019ctu}.

Events of interest are selected using a two-tiered trigger system. The first level, 
composed of custom hardware processors, uses information from the calorimeters and muon detectors to select 
events at a rate of around 100\unit{kHz} within a fixed latency of about 4\mus~\cite{CMS:2020cmk}. 
The second level, known as the high-level trigger, consists of a farm of processors running a version of the 
full event reconstruction software optimized for fast processing, and reduces the event rate to 
around 1\unit{kHz} before data storage~\cite{CMS:2016ngn}.
Events are selected using a combination of single-, double-, and triple-lepton triggers. The minimum lepton transverse momentum (\pt) requirements are chosen to ensure full trigger efficiency for the offline selection. Trigger efficiencies, measured with independent \ptmiss  triggers, exceed 95\% over the full \pt spectrum, with combined statistical and systematic uncertainties below 2\%~\cite{CMS:2023xyc}.

Monte Carlo (MC) simulated events are used for  modelling  most of the background processes and the FCNC signals.
The signal samples for the production ($\Pp \Pp \to \PQt \Pell \PAell$) and decay ($\Pp \Pp \to \ttbar \to \Pell \PAell \PQq \PAQt$) modes are generated at leading order (LO) in quantum chromodynamics (QCD) with the \MGvATNLO 2.6.0 program~\cite{MadGraph5_aMCatNLO,Frederix:2012ps,Alwall:2007fs}. Signal samples involving the Higgs boson are generated separately. Samples with $\PQt\PQu$ and $\PQt\PQc$ FCNCs are also generated independently, since no interference is expected between these two processes.
The EFT effects are included  using the \textsc{dim6top} Universal FeynRules Output model~\cite{AguilarSaavedra:2018nen}. To evaluate EFT predictions at arbitrary points in the WC parameter space, the \MGvATNLO reweighting technique is employed~\cite{CMS:2023xyc,CMS:2020lrr}. Since the FCNC cross section is a quadratic polynomial in the WCs, a finite
 set of weights is sufficient to reconstruct the cross section in any
 phase-space region or observable bin~\cite{Belvedere:2024nzh}. The event weights in the decay mode are corrected to account for higher-order effects, ensuring consistency with the inclusive \ttbar cross section~\cite{CMS:2024dzv}.

Backgrounds from $\ttW$, $\ttZ$, $\ttH$, $\tZq$, diboson, and triboson are modelled using next-to-LO (NLO) matrix elements with \MGvATNLO~\cite{MadGraph5_aMCatNLO,Frederix:2012ps,Alwall:2007fs} and \POWHEG v2~\cite{POWHEG1,POWHEG2,POWHEG3,POWHEG_VV,Frixione:2007nw}.  In addition, dedicated simulated samples of photon-associated processes, such as $\ttG$, $\WG$, $\ZG$, and $\tqG$, are included as separate background contributions to improve the modelling of events containing photon conversions.

Both the signal and background samples are interfaced with \PYTHIA~8.205~\cite{Sjostrand:2014zea} for parton showering, fragmentation, and hadronization. 
The underlying event is modeled using \PYTHIA with the CP5 tune~\cite{Sirunyan:2019dfx}. 
Event generation employs next-to-NLO parton distribution functions (PDFs) from the NNPDF3.1 set~\cite{Ball:2017nwa}. 
All MC samples are processed through the full CMS detector simulation based on \GEANTfour~\cite{Agostinelli:2002hh}. 
Simulated samples include additional pp interactions within
the same or nearby bunch crossings (pileup) and are reweighted to reproduce the distribution of the mean number of interactions observed in data.

The FCNC signal events are characterized by the presence of multiple leptons and jets, with exactly one jet identified as a \PQb-tagged jet. 
Electron, muon, jet, and \PQb-tagged jet reconstruction, identification, and selection criteria follow those used in the similar CMS multilepton analysis described in Ref.~\cite{CMS:2023xyc}. 
Each event is required to contain either two same-sign charged leptons (\twolss channel) or three leptons (\threel channel). 
In the \twolss channel, the leading (subleading) lepton is required to satisfy $\pt>25~(15)\GeV$.  Electrons (muons) are required to lie within $\abs{\eta}<2.5~(2.4)$.
In the \threel channel, the three leptons must satisfy $\pt > 25$, 15, and 10\GeV, respectively, 
with the same $\abs{\eta}$ requirements as in the \twolss; if the third lepton is an electron, 
a tighter threshold of $\pt > 15\GeV$ is applied.

Although it is not possible to fully isolate individual processes for each WC, the data set is divided into subsamples based on specific signal features to maximize the statistical power for probing different processes and WCs. 
Signal events associated with the Higgs boson (\ctp) predominantly populate the \twolss channel, mainly through the $\PH \to \PW\PW \to \Pell\PGn\PQq\PQq$ decay chain. 
Processes involving the $\PZ$ boson (\cpQM, \cpt, and \ctZ) mainly contribute to the \threel channel, where an opposite-sign same-flavor (OSSF) lepton pair has an invariant mass ($m_{\Pell\Pell}$) consistent with that of the $\PZ$ boson mass ($m_{\PZ}$). 
In contrast, two-quark--two-lepton operators, with WCs $\ctl$, $\cte$, $\cQlM$, $\cQe$, $\ctlS$, and $\ctlT$,  mainly contribute to the \threel channel, where the OSSF lepton-pair invariant mass lies far from the $\PZ$ boson mass.
Signal events arising from decay topologies typically contain one additional light quark at the matrix-element level compared to production processes. 

To exploit these features, events with three leptons are required to contain at least one OSSF lepton pair. A mass window of 10\GeV around the $\PZ$ boson mass is used to identify events consistent with a $\PZ$ boson decay. Events containing an OSSF lepton pair satisfying $\abs{m_{\Pell\Pell}-m_{\PZ}}<10\GeV$ are assigned to the ``\threelonZ'' category, while the remaining events are assigned to the ``\threeloffZ'' category. 
Events with a dilepton invariant mass below 12\GeV are rejected to suppress backgrounds from low-mass resonances. In addition, events in the \threeloffZ category are required to satisfy 
$\abs{m_{3\Pell}-m_{\PZ}} > 10\GeV$ in order to suppress Drell--Yan background with an associated final-state radiation photon conversion peaking at the \PZ boson mass.

Selected events in the \twolss and \threel channels are further subdivided according to jet and \PQb-jet multiplicity, where jets are required to satisfy $\pt > 30\GeV$ and $\abs{\eta}<2.4$. The \PQb-tagged jets are identified using the \textsc{DeepJet} algorithm~\cite{Bols:2020bkb,CMS-DP-2023-005,Cacciari:2011ma}. A medium working point of the \PQb tagging algorithm is used, providing a \PQb-jet identification efficiency of about 70\% and a misidentification rate of approximately 1\% for light-flavor quark and gluon jets. 

The FCNC signal events contain exactly one \PQb jet, whereas SM background events can contain more than one \PQb jet. Events are therefore categorized according to the \PQb jet multiplicity into a region with exactly one \PQb jet and a region with more than one \PQb jet, referred to as the signal region (SR) and control region (CR), respectively.
To further exploit differences in jet multiplicity between FCNC signal events produced via the production and decay modes, events in the SR are subdivided according to the jet multiplicity. In the \twolss channel, the jet multiplicity threshold is optimized using the Higgs FCNC signal in the $\PH \to \PW\PW \to \Pell\PGn\PQq\PQq$ decay mode. No additional categorization based on jet multiplicity is applied in the CR.
A summary of the SR and CR selection criteria is provided in Table~\ref{tab:selection}.

\begin{table*}[htbp]
\centering
\caption{Summary of the SR and CR categorization requirements.}
\label{tab:selection}

\renewcommand{\arraystretch}{1.15}

\begin{scotch}{lccc}
 & \twolss & \threelonZ & \threeloffZ \\
\hline

\multicolumn{4}{c}{{Lepton requirements}} \\
\hline

Leptons
& 2 & 3 & 3 \\

$m_{\Pell\Pell}$ 
& $>$12\GeV
& $>$12\GeV
& $>$12\GeV \\

OSSF pair
& --
& $\abs{m_{\Pell\Pell}-m_{\PZ}}<10\GeV$
& $\abs{m_{\Pell\Pell}-m_{\PZ}}>10\GeV$ \\

$m_{\Pell\Pell\Pell}$
& --
& --
& $\abs{m_{\Pell\Pell\Pell}-m_{\PZ}}>10\GeV$ \\

\hline
\multicolumn{4}{c}{{Event categorization}} \\
\hline

\PQb jets
& \multicolumn{3}{c}{SR: $=1$, \quad CR: $>1$} \\

SR jet categories
& $\leq$3, $>$3
& $=$1, $>$1
& $=$1, $>$1 \\

CR jet categories
& \multicolumn{3}{c}{No additional categorization} \\

\hline
\multicolumn{4}{c}{{Sensitive Wilson coefficients}} \\
\hline
\multirow{2}{*}{Sensitive WCs} & \multirow{2}{*}{\ctp} & \multirow{2}{*}{\cpQM, \cpt, \ctZ} & \ctl, \cte, \cQlM, \\
                               &                       &                                    & \cQe, \ctlS, \ctlT \\
\end{scotch}
\end{table*}

Although all WCs contribute to the \twolss channel, this channel is primarily optimized for \ctp using a binary boosted decision tree (BDT) classifier~\cite{Hoecker:2007ht}. The highest sensitivity for the remaining coefficients is found in the \threel channels.
As summarized in Table~\ref{tab:selection}, several coefficients contribute significantly to the \threelonZ and \threeloffZ categories. Accordingly, multiclass BDT classifiers are trained for the \threel categories. Studies based on MC simulation show that the kinematic distributions of the FCNC signals are very similar for nonzero \cpQM and \cpt in the \threelonZ channel, despite the different Lorentz structures of the corresponding tqZ interactions. A similar behavior is observed for the vector-like four-fermion operators \ctl, \cte, \cQlM, and \cQe in the \threeloffZ channel, which differ mainly in their chiral structure. Since the primary goal of this analysis is to maximize sensitivity to FCNC signals from different EFT sources relative to the SM backgrounds, rather than to distinguish between closely related Lorentz or chiral structures, signal events within each of these groups are combined in the BDT training. This approach improves the statistical robustness and sensitivity of the classifiers while retaining sensitivity to the targeted FCNC production modes. Owing to the limited sensitivity to the \ctG and \ctA operators in the multilepton final state, their distinct kinematic features are not explicitly targeted in the BDT training.

For the BDT training, all simulated background samples are normalized to their theoretical cross sections. Signal events are assigned weights corresponding to the chosen nonzero WCs, while the same underlying simulated events are used for all EFT scenarios. Consequently, the EFT information is encoded in the relative event weights. 

The output of the BDT classifier is used to construct a discriminant 
sensitive to the presence of FCNC processes. 
In the multiclass setup, the BDT assigns each event a set of normalized 
outputs that can be interpreted as class probability, 
$P(\text{WC}_i)$ for the signal classes associated with different 
WCs and $P_{\text{BG}}$ for the background class. 
We define the ``final discriminant'' as
\begin{equation}
\frac{\sum_i P(\text{WC}_i)}{P_{\text{BG}}},
\end{equation}
which combines the information from all 
signal hypotheses. 
Here, the sum runs over all considered FCNC signal classes. 
In the \twolss channel, the BDT output is used as the final discriminant.

To enhance the sensitivity of the BDT classifier, a set of input features is constructed from a full event reconstruction performed with the jigsaw algorithm~\cite{Jackson:2017gcy}, which reconstructs the top quark and the associated Higgs or $\PZ$ boson.
The jigsaw algorithm reconstructs the four-momenta of intermediate particles in events with missing momentum by assigning reconstructed objects to hypothesised decay chains and solving for the best kinematic configuration consistent with the event topology and momentum conservation.
The reconstruction assumes the $\PH \to \PW\PW$ decay in the \twolss channel. 
By relying only on the reconstructed four-momenta of the  top quark with its SM decay and the Higgs or \PZ boson, the method is valid for the FCNC signal events in the production and decay modes. 
In the \threel channel, the $\PZ$ boson candidate is defined as the OSSF lepton pair with the highest invariant mass. For events with more than one \PQb-tagged jet, the leading-\pt \PQb-tagged jet is used in the jigsaw reconstruction.

The following kinematic variables are used as BDT inputs: 
the \pt of the selected leptons, leading jet, and \PQb jet; 
the total number of jets; 
the difference in azimuthal angle between the leading ($\Pell_1$) and subleading ($\Pell_2$) leptons, $\Delta\phi(\Pell_1,\Pell_2)$; 
their angular separation, $\Delta R(\Pell_1,\Pell_2) = \sqrt{\smash[b]{(\Delta\eta)^2 + (\Delta\phi)^2}}$; 
and the \pt of the $\Pell_1\Pell_2$ system.

Additionally, the jigsaw-reconstructed top quark kinematic variables, including its \pt, $\eta$, mass, and $\Delta R$ between the $\PW$ boson from the top quark decay and the \PQb jet, are included. 
In the \twolss categories, the jigsaw-reconstructed Higgs boson \pt, $\eta$, mass, the reconstructed masses of $\PW_1$ and $\PW_2$ from the Higgs boson decay, and $\Delta R( \PQb\; \text{jet}_1, \PQb\; \text{jet}_2)$ are used. 
In the \threel categories, the reconstructed $\PZ$ boson \pt, $\eta$, and mass are considered.

The SR and CR, defined in Table~\ref{tab:selection}, receive contributions from both irreducible and reducible SM backgrounds. 
Irreducible backgrounds arise from processes with only genuine prompt leptons, including $\ttW$, $\ttZ$, $\ttH$, $\tZq$, diboson, and triboson production, and are estimated from simulation. 
Reducible backgrounds originate from nonprompt leptons, charge misidentification, and photon conversions, and their estimation strategy follows Ref.~\cite{CMS:2023xyc}. 
The dominant reducible contribution comes from nonprompt leptons, primarily from heavy-flavor decays, and is estimated directly from collision data using the misidentification probability method~\cite{CMS:2020mpn} in CRs with relaxed lepton identification. Events are weighted by probabilities measured separately for electrons and muons in multijet-enriched data as functions of lepton $\pt$ and $\eta$.
Charge misidentification, which mainly affects electrons in \twolss events from \ttbar production, is estimated using data with opposite-sign electron pairs, weighted by the misidentification rate measured in $\PZ \to \Pe\Pe$ simulation.
Photon conversions are suppressed through electron identification requirements on inner-track hits and conversion vetoes, with the remaining contribution estimated from simulation~\cite{CMS:2017moi}.

The systematic uncertainties considered in this analysis are largely consistent with those reported in Ref.~\cite{CMS:2023xyc}. 
Experimental sources include the integrated luminosity, lepton identification and isolation efficiencies, trigger efficiency, pileup modelling, jet energy scale and resolution, \PQb jet identification efficiency, the  level-1 ECAL trigger inefficiency~\cite{CMS:2020cmk}, and the estimation of misidentified-lepton  and charge misreconstruction backgrounds from control samples in data, all treated following the same methodology. 
Theoretical sources are also handled as in Ref.~\cite{CMS:2023xyc}, and include the choice of PDFs and QCD coupling, variations in renormalization and factorization scales, and parton shower modelling. 
Normalization uncertainties are further assigned to the cross sections of the main background processes: 10\% for triboson, 6\% for diboson, 10\% for $\tZq$, 5\% for photon conversions, and 15\% for \ttZ, \ttH, and \tWZ~\cite{CMS:2022tkv}.
Owing to the discrepancy between the theoretical prediction and the measured \ttW cross section in a similar phase space but requiring two \PQb-tagged jets, a 50\% uncertainty is assigned to the \ttW production cross section~\cite{CMS:2022tkv,ATLAS:2024moy}.
In this analysis, the \ptvecmiss is used in the jigsaw reconstruction~\cite{CMS:2019ctu}. Therefore, the uncertainty associated with unclustered energy (energy deposits not assigned to reconstructed physics objects) is included in the \ptvecmiss measurement. In addition, the effects of the jet energy scale and jet energy resolution uncertainties are propagated to \ptvecmiss.

\begin{figure*}[tbh!]
      \includegraphics[width=1\textwidth]{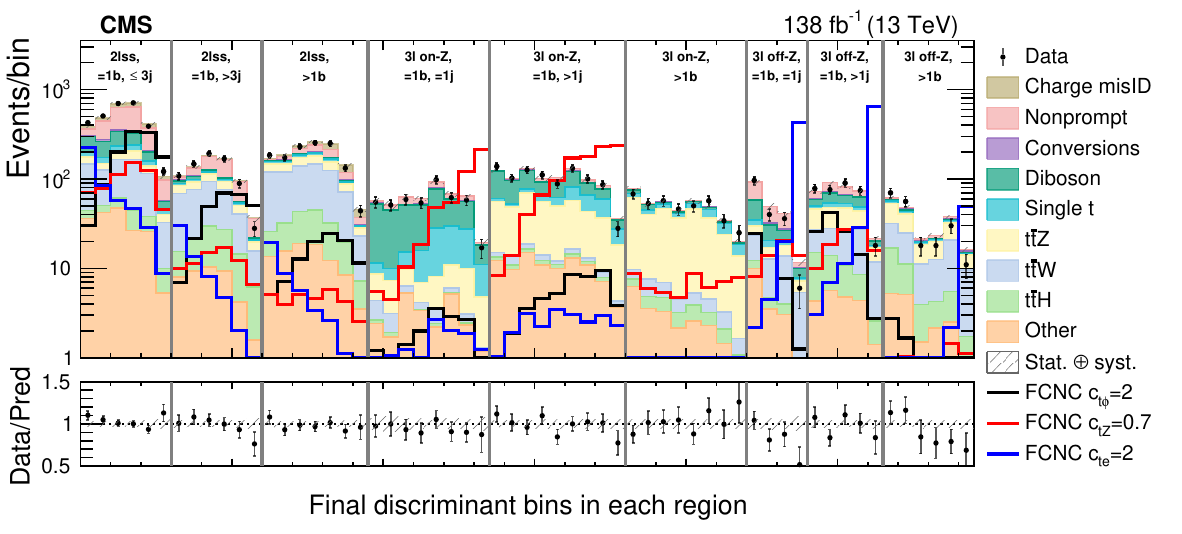}
  \caption{Background-only post-fit distributions of the final discriminant are shown for the $\twolss$, $\threelonZ$, and $\threeloffZ$ channels in the SRs and CRs. Signal predictions for three representative WCs ($\ctp$, $\ctZ$, and $\cte$) are overlaid for illustration.  The lower panels show the ratio of the data to the total SM background, with the total uncertainties indicated by the shaded gray band.}
\label{fig:finalPlotTU}
\end{figure*}

A maximum likelihood fit is performed simultaneously across three main categories (\twolss, \threelonZ, and \threeloffZ), each further divided into three subcategories based on jet and \PQb-tagged jet multiplicities (SR and CR), and across the different data-taking periods, using the \textsc{combine}~\cite{CMS:2024onh} tool. In each subcategory and data-taking period, the distribution of the final discriminant, constructed from the BDT outputs, is used. The expected signal yield in each bin is parametrized as a quadratic function of the 24 WCs, which serve as the parameters of interest in the fit. Signal and background yields also incorporate the impact of systematic uncertainties, which are included as nuisance parameters in the likelihood.  Figure~\ref{fig:finalPlotTU} shows the post-fit distributions for the background-only fit across all included bins. The fitted background prediction is in good agreement with the observed data within the associated uncertainties. In addition, the signal predictions for $\PQt\PQu$ FCNC in the production mode are shown for three representative WCs, $(\ctp, \ctZ,$ and $\cte)$, for illustration.

\begin{figure*}[tbh!]
  \centering
      \includegraphics[width=0.85\textwidth]{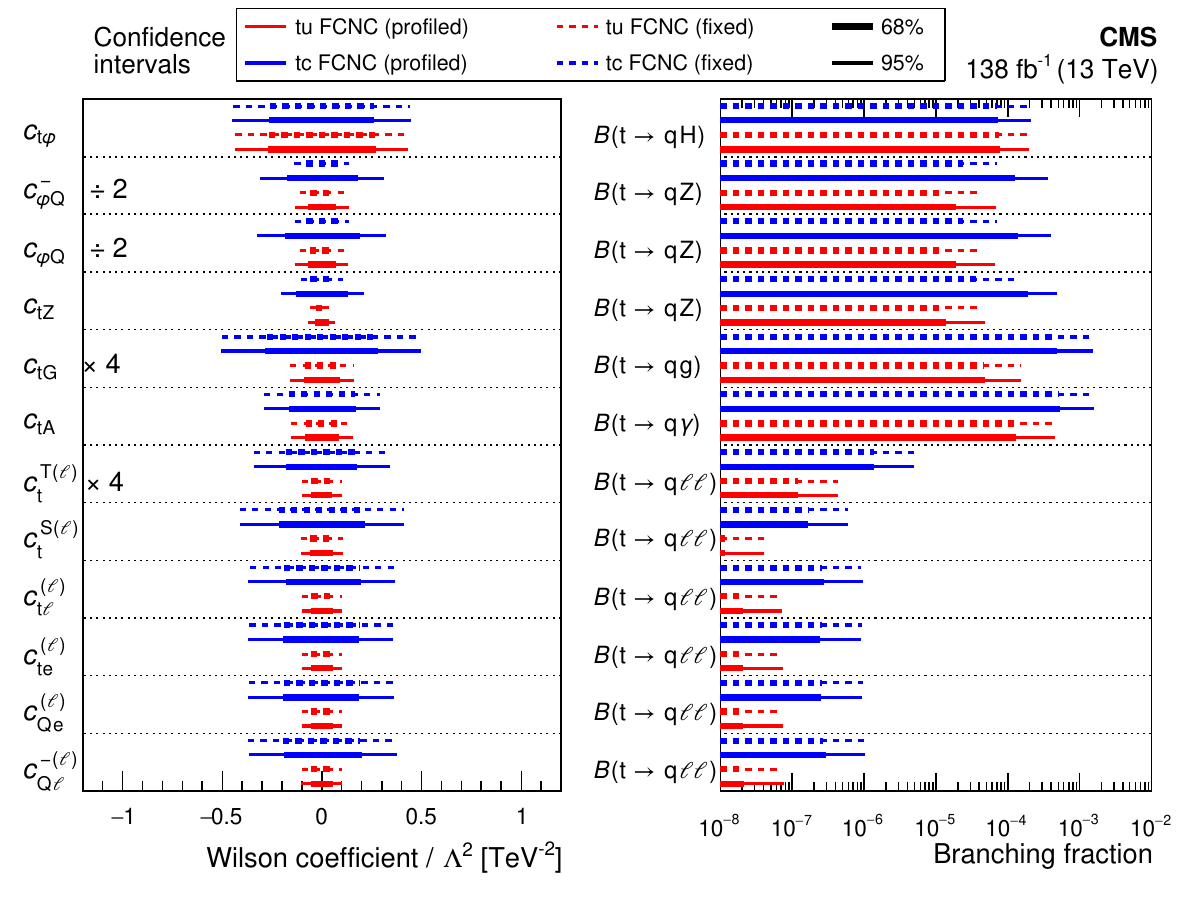}
  \caption{The $68\%$ (thick lines) and $95\%$ (thin lines) confidence intervals for the $\PQt\PQu$ and $\PQt\PQc$ FCNC WCs, obtained from profiled (solid lines) and fixed (dashed lines) fits. The limits on the WCs (left) are translated into the 
corresponding limits on the branching fractions (right).}
\label{fig:limitsummary}
\end{figure*}

Confidence intervals at 68 and 95\% are obtained for each WC by performing likelihood scans, profiling the remaining 23 WCs and all nuisance parameters (profiled). In addition, scans are repeated with the other 23 WCs fixed to zero (fixed) to assess individual sensitivities. 
The resulting limits on the WCs are translated into constraints on the corresponding top quark FCNC branching fractions. 
For each WC, the partial decay width of the corresponding FCNC top quark decay is calculated with all other WCs set to zero, and the branching fraction is obtained as the ratio of this partial width to the total top quark decay width.
Figure~\ref{fig:limitsummary} shows the 68 and 95\% intervals for each WC and the associated branching fractions.  Tabulated results are provided in the HEPData record for this analysis~\cite{hepdata}.
Using MC pseudo-experiments, we verified that the $68$ and $95\%$ confidence intervals obtained using Wilks theorem~\cite{Wilks:1938dza} are consistent with the exact frequentist intervals derived from pseudo-experiment distributions, and are in some cases slightly more conservative~\cite{Bernlochner:2022oiw}. This demonstrates that the asymptotic approximation provides reliable confidence limits throughout the investigated parameter space.

In summary, a search for flavor-changing neutral current (FCNC) interactions involving the top quark is performed using proton--proton collision data at $\sqrt{s}=13\TeV$ recorded by the CMS experiment in 2016--2018, corresponding to an integrated luminosity of 138\fbinv. Events with same-sign charged dilepton or trilepton final states and exactly one \PQb-tagged jet are selected, and boosted decision trees are used to enhance the separation of signal from background. A binned maximum likelihood fit to the discriminant distributions is used to extract constraints on the underlying effective couplings. No significant deviation from the standard model (SM) prediction is observed. Confidence intervals at 68 and 95\% confidence level are set on 24 Wilson coefficients. These results constitute the first simultaneous constraints on the complete set of relevant top quark FCNC operators in the SM effective field theory (EFT) and establish a global benchmark for future top quark FCNC measurements and their combination at the LHC. In addition, they provide the first direct experimental constraints on FCNC processes mediated by two-quark--two-lepton operators.
While a direct comparison is inherently dependent on the choice of parameterization 
and operator definitions, the upper limits placed on the $\mathcal{B}(\PQt \to \PH\Pq)$ 
and $\mathcal{B}(\PQt \to \PZ\Pq)$ branching fractions within the EFT framework are 
found to be similar or more restrictive than prior results based on anomalous couplings~\cite{ATLAS:2024mih,CMS:2024ubt,ATLAS:2023qzr,CMS:2017wcz}.

\begin{acknowledgments}
We congratulate our colleagues in the CERN accelerator departments for the excellent performance of the LHC and thank the technical and administrative staffs at CERN and at other CMS institutes for their contributions to the success of the CMS effort. In addition, we gratefully acknowledge the computing centers and personnel of the Worldwide LHC Computing Grid and other centers for delivering so effectively the computing infrastructure essential to our analyses. Finally, we acknowledge the enduring support for the construction and operation of the LHC, the CMS detector, and the supporting computing infrastructure provided by the following funding agencies: SC (Armenia), BMFWF and FWF (Austria); FNRS and FWO (Belgium); CNPq, CAPES, FAPERJ, FAPERGS, and FAPESP (Brazil); MES and BNSF (Bulgaria); CERN; CAS, MoST, and NSFC (China); MINCIENCIAS (Colombia); MSES and CSF (Croatia); RIF (Cyprus); SENESCYT (Ecuador); ERC PRG and PSG, TARISTU24-TK10 and MoER TK202 (Estonia); Academy of Finland, MEC, and HIP (Finland); CEA and CNRS/IN2P3 (France); SRNSF (Georgia); BMFTR, DFG, and HGF (Germany); GSRI (Greece); MATE and NKFIH (Hungary); DAE and DST (India); IPM (Iran); SFI (Ireland); INFN (Italy); MSIT and NRF (Republic of Korea); MES (Latvia); LMTLT (Lithuania); MOE and UM (Malaysia); BUAP, CINVESTAV, CONACYT, LNS, SEP, and UASLP-FAI (Mexico); MOS (Montenegro); MBIE (New Zealand); PAEC (Pakistan); MSHE, NSC, and NAWA (Poland); FCT (Portugal);  MESTD (Serbia); MICIU/AEI and PCTI (Spain); MOSTR (Sri Lanka); Swiss Funding Agencies (Switzerland); MST (Taipei); MHESI (Thailand); TUBITAK and TENMAK (T\"{u}rkiye); NASU (Ukraine); STFC (United Kingdom); DOE and NSF (USA).
\end{acknowledgments}\section*{Data availability} Release and preservation of data used by the CMS Collaboration as the basis for publications is guided by the  \href{https://doi.org/10.7483/OPENDATA.CMS.1BNU.8V1W}{CMS data preservation, re-use and open access policy}.

\bibliography{auto_generated} 

@ARTICLE{Glashow:1970gm,
	AUTHOR=	"Glashow, S. L. and Iliopoulos, J. and Maiani, L.",
	TITLE=	"Weak interactions with lepton--hadron symmetry",
	JOURNAL=	"Phys. Rev. D",
	VOLUME=	"2",
	PAGES=	"1285",
	YEAR=	"1970",
	DOI=	"10.1103/PhysRevD.2.1285",
}

@ARTICLE{AguilarSaavedra:2004wm,
	AUTHOR=	"Aguilar-Saavedra, J. A.",
	TITLE=	"{Top flavor-changing neutral interactions: theoretical expectations and experimental detection}",
	BOOKTITLE=	"{Particle physics phenomenology at high energy colliders. Proceedings, Final meeting of the European Union Network, Montpellier, France, September 26-27, 2004}",
	JOURNAL=	"Acta Phys. Polon. B",
	VOLUME=	"35",
	YEAR=	"2004",
	PAGES=	"2695",
	EPRINT=	"hep-ph/0409342",
	ARCHIVEPREFIX=	"arXiv",
	SLACCITATION=	"%%CITATION = HEP-PH/0409342;%%",
}

@ARTICLE{Yang:1997dk,
	AUTHOR=	"Yang, Jin Min and Young, Bing-Lin and Zhang, X.",
	TITLE=	"{Flavor-changing top quark decays in R-parity-violating supersymmetric models}",
	JOURNAL=	"Phys. Rev. D",
	VOLUME=	"58",
	YEAR=	"1998",
	PAGES=	"055001",
	DOI=	"10.1103/PhysRevD.58.055001",
	EPRINT=	"hep-ph/9705341",
	ARCHIVEPREFIX=	"arXiv",
	REPORTNUMBER=	"AMES-HET-97-5",
	SLACCITATION=	"%%CITATION = HEP-PH/9705341;%%",
}

@ARTICLE{Delepine:2004hr,
	AUTHOR=	"Delepine, D. and Khalil, S.",
	TITLE=	"{Top} flavour violating decays in general supersymmetric models",
	EPRINT=	"hep-ph/0406264",
	ARCHIVEPREFIX=	"arXiv",
	REPORTNUMBER=	"IPPP-04-05, DCPT-04-10",
	DOI=	"10.1016/j.physletb.2004.08.025",
	JOURNAL=	"Phys. Lett. B",
	VOLUME=	"599",
	PAGES=	"62",
	YEAR=	"2004",
}

@ARTICLE{Cao:2007dk,
	AUTHOR=	"Cao, J. J. and Eilam, G. and Frank, M. and Hikasa, K. and Liu, G. L. and Turan, I. and Yang, J. M.",
	TITLE=	"{Supersymmetry-induced} flavor-changing neutral-current top-quark processes at the {CERN Large Hadron Collider}",
	EPRINT=	"hep-ph/0702264",
	ARCHIVEPREFIX=	"arXiv",
	REPORTNUMBER=	"CUMQ-HEP-144",
	DOI=	"10.1103/PhysRevD.75.075021",
	JOURNAL=	"Phys. Rev. D",
	VOLUME=	"75",
	PAGES=	"075021",
	YEAR=	"2007",
}

@ARTICLE{Yang:2018utw,
	AUTHOR=	"Yang, Jin-Lei and Feng, Tai-Fu and Zhang, Hai-Bin and Ning, Guo-Zhu and Yang, Xiu-Yi",
	TITLE=	"{Top} quark decays with flavor violation in the {B-LSSM}",
	EPRINT=	"1806.01476",
	ARCHIVEPREFIX=	"arXiv",
	PRIMARYCLASS=	"hep-ph",
	DOI=	"10.1140/epjc/s10052-018-5919-5",
	JOURNAL=	"Eur. Phys. J. C",
	VOLUME=	"78",
	PAGES=	"438",
	YEAR=	"2018",
}

@ARTICLE{Atwood:1996vj,
	AUTHOR=	"Atwood, David and Reina, Laura and Soni, Amarjit",
	TITLE=	"{Phenomenology of two Higgs doublet models with flavor-changing neutral currents}",
	JOURNAL=	"Phys. Rev. D",
	VOLUME=	"55",
	YEAR=	"1997",
	PAGES=	"3156",
	DOI=	"10.1103/PhysRevD.55.3156",
	EPRINT=	"hep-ph/9609279",
	ARCHIVEPREFIX=	"arXiv",
	REPORTNUMBER=	"JL-TH-96-15, JLAB-THY-96-08, CEBAF-PREPRINT-JL-TH-96-15",
	SLACCITATION=	"%%CITATION = HEP-PH/9609279;%%",
}

@ARTICLE{Bejar:2000ub,
	AUTHOR=	"Bejar, Santi and Guasch, Jaume and Sola, Joan",
	TITLE=	"{Loop} induced flavor changing neutral decays of the top quark in a general {two-Higgs-doublet} model",
	EPRINT=	"hep-ph/0011091",
	ARCHIVEPREFIX=	"arXiv",
	REPORTNUMBER=	"UAB-FT-491, KA-TP-22-2000",
	DOI=	"10.1016/S0550-3213(01)00044-X",
	JOURNAL=	"Nucl. Phys. B",
	VOLUME=	"600",
	PAGES=	"21",
	YEAR=	"2001",
}

@ARTICLE{ATLAS:2024mih,
	AUTHOR=	"{ATLAS Collaboration}",
	TITLE=	"{Search} for flavour-changing neutral-current couplings between the top quark and the Higgs boson in multi-lepton final states in 13~TeV pp collisions with the {ATLAS} detector",
	EPRINT=	"2404.02123",
	ARCHIVEPREFIX=	"arXiv",
	PRIMARYCLASS=	"hep-ex",
	REPORTNUMBER=	"CERN-EP-2024-070, CERN-EP-2024-070",
	DOI=	"10.1140/epjc/s10052-024-12994-1",
	JOURNAL=	"Eur. Phys. J. C",
	VOLUME=	"84",
	PAGES=	"757",
	YEAR=	"2024",
}

@ARTICLE{CMS:2024ubt,
	AUTHOR=	"{CMS Collaboration}",
	TITLE=	"{Search} for flavor-changing neutral current interactions of the top quark mediated by a {Higgs} boson in proton-proton collisions at 13 {TeV}",
	EPRINT=	"2407.15172",
	ARCHIVEPREFIX=	"arXiv",
	PRIMARYCLASS=	"hep-ex",
	REPORTNUMBER=	"CMS-TOP-22-002, CERN-EP-2024-174",
	DOI=	"10.1103/95q6-vvlp",
	JOURNAL=	"Phys. Rev. D",
	VOLUME=	"112",
	PAGES=	"032008",
	YEAR=	"2025",
}

@ARTICLE{ATLAS:2021amo,
	AUTHOR=	"{ATLAS Collaboration}",
	TITLE=	"{Search for flavour-changing neutral-current interactions of a top quark and a gluon in pp collisions at $\sqrt{s}=13$~TeV with the ATLAS detector}",
	EPRINT=	"2112.01302",
	ARCHIVEPREFIX=	"arXiv",
	PRIMARYCLASS=	"hep-ex",
	REPORTNUMBER=	"CERN-EP-2021-234",
	DOI=	"10.1140/epjc/s10052-022-10182-7",
	JOURNAL=	"Eur. Phys. J. C",
	VOLUME=	"82",
	PAGES=	"334",
	YEAR=	"2022",
}

@ARTICLE{CMS:2016uzc,
	AUTHOR=	"{CMS Collaboration}",
	TITLE=	"{Search for anomalous Wtb couplings and flavour-changing neutral currents in $t$-channel single top quark production in pp collisions at $\sqrt{s} =$ 7 and 8 TeV}",
	EPRINT=	"1610.03545",
	ARCHIVEPREFIX=	"arXiv",
	PRIMARYCLASS=	"hep-ex",
	REPORTNUMBER=	"CMS-TOP-14-007, CERN-EP-2016-207",
	DOI=	"10.1007/JHEP02(2017)028",
	JOURNAL=	"JHEP",
	VOLUME=	"02",
	PAGES=	"028",
	YEAR=	"2017",
}

@ARTICLE{ATLAS:2022per,
	AUTHOR=	"{ATLAS Collaboration}",
	TITLE=	"{Search for flavour-changing neutral-current couplings between the top quark and the photon with the ATLAS detector at $\sqrt{s}=13$~TeV}",
	EPRINT=	"2205.02537",
	ARCHIVEPREFIX=	"arXiv",
	PRIMARYCLASS=	"hep-ex",
	REPORTNUMBER=	"CERN-EP-2022-042",
	DOI=	"10.1016/j.physletb.2022.137379",
	JOURNAL=	"Phys. Lett. B",
	VOLUME=	"842",
	PAGES=	"137379",
	YEAR=	"2023",
	NOTE=	"[Erratum: \href{https://doi.org/10.1016/j.physletb.2023.138286}{Phys. Lett. B 847 (2024) 138286}]",
}

@ARTICLE{CMS:2023bjm,
	AUTHOR=	"{CMS Collaboration}",
	TITLE=	"{Search for flavor changing neutral current interactions of the top quark in final states with a photon and additional jets in proton-proton collisions at $\sqrt{s}=13$~TeV}",
	EPRINT=	"2312.08229",
	ARCHIVEPREFIX=	"arXiv",
	PRIMARYCLASS=	"hep-ex",
	REPORTNUMBER=	"CMS-TOP-21-013, CERN-EP-2023-214",
	DOI=	"10.1103/PhysRevD.109.072004",
	JOURNAL=	"Phys. Rev. D",
	VOLUME=	"109",
	PAGES=	"072004",
	YEAR=	"2024",
}

@ARTICLE{ATLAS:2023qzr,
	AUTHOR=	"{ATLAS Collaboration}",
	TITLE=	"{Search for flavor-changing neutral-current couplings between the top quark and the Z boson with proton-proton collisions at $\sqrt{s}=13$~TeV with the ATLAS detector}",
	EPRINT=	"2301.11605",
	ARCHIVEPREFIX=	"arXiv",
	PRIMARYCLASS=	"hep-ex",
	REPORTNUMBER=	"CERN-EP-2022-044",
	DOI=	"10.1103/PhysRevD.108.032019",
	JOURNAL=	"Phys. Rev. D",
	VOLUME=	"108",
	PAGES=	"032019",
	YEAR=	"2023",
}

@ARTICLE{CMS:2017wcz,
	AUTHOR=	"{CMS Collaboration}",
	TITLE=	"{Search for associated production of a Z boson with a single top quark and for tZ flavour-changing interactions in pp collisions at $ \sqrt{s}=8 $ TeV}",
	EPRINT=	"1702.01404",
	ARCHIVEPREFIX=	"arXiv",
	PRIMARYCLASS=	"hep-ex",
	REPORTNUMBER=	"CMS-TOP-12-039, CERN-EP-2016-324",
	DOI=	"10.1007/JHEP07(2017)003",
	JOURNAL=	"JHEP",
	VOLUME=	"07",
	PAGES=	"003",
	YEAR=	"2017",
}

@UNPUBLISHED{AguilarSaavedra:2018nen,
	AUTHOR=	"Barducci, D. and others",
	EDITOR=	"Aguilar-Saavedra, Juan Antonio and Degrande, C. and Durieux, G. and Maltoni, F. and Vryonidou, E. and Zhang, C.",
	TITLE=	"{Interpreting top-quark LHC measurements in the standard-model effective field theory}",
	YEAR=	"2018",
	EPRINT=	"1802.07237",
	ARCHIVEPREFIX=	"arXiv",
	PRIMARYCLASS=	"hep-ph",
	REPORTNUMBER=	"CERN-LPCC-2018-01",
	SLACCITATION=	"%%CITATION = ARXIV:1802.07237;%%",
}

@ARTICLE{CMS:2023xyc,
	AUTHOR=	"{CMS Collaboration}",
	TITLE=	"{Search} for physics beyond the standard model in top quark production with additional leptons in the context of effective field theory",
	EPRINT=	"2307.15761",
	ARCHIVEPREFIX=	"arXiv",
	PRIMARYCLASS=	"hep-ex",
	REPORTNUMBER=	"CMS-TOP-22-006, CERN-EP-2023-124",
	DOI=	"10.1007/JHEP12(2023)068",
	JOURNAL=	"JHEP",
	VOLUME=	"12",
	PAGES=	"068",
	YEAR=	"2023",
}

@ARTICLE{ATLAS:2025yww,
	AUTHOR=	"{ATLAS Collaboration}",
	TITLE=	"{Measurement of high-mass $t\bar{t}\ell^{+}\ell^{-}$ production and lepton flavour universality-inspired effective field theory interpretations at $\sqrt{s}=13$ TeV with the ATLAS detector}",
	EPRINT=	"2504.05919",
	ARCHIVEPREFIX=	"arXiv",
	PRIMARYCLASS=	"hep-ex",
	REPORTNUMBER=	"CERN-EP-2025-053",
	DOI=	"10.1140/epjc/s10052-025-14695-9",
	JOURNAL=	"Eur. Phys. J. C",
	VOLUME=	"85",
	PAGES=	"1434",
	YEAR=	"2025",
}

@ARTICLE{CMS:2025ugn,
	AUTHOR=	"{CMS Collaboration}",
	TITLE=	"{Combined} effective field theory interpretation of {Higgs} boson, electroweak vector boson, top quark, and multijet measurements",
	EPRINT=	"2504.02958",
	ARCHIVEPREFIX=	"arXiv",
	PRIMARYCLASS=	"hep-ex",
	REPORTNUMBER=	"CMS-SMP-24-003, CERN-EP-2025-035",
	DOI=	"10.1140/epjc/s10052-025-14997-y",
	JOURNAL=	"Eur. Phys. J. C",
	VOLUME=	"86",
	PAGES=	"331",
	YEAR=	"2026",
}

@ARTICLE{Durieux:2014xla,
	AUTHOR=	"Durieux, Gauthier and Maltoni, Fabio and Zhang, Cen",
	TITLE=	"{Global} approach to top-quark flavor-changing interactions",
	EPRINT=	"1412.7166",
	ARCHIVEPREFIX=	"arXiv",
	PRIMARYCLASS=	"hep-ph",
	REPORTNUMBER=	"CP3-14-85",
	DOI=	"10.1103/PhysRevD.91.074017",
	JOURNAL=	"Phys. Rev. D",
	VOLUME=	"91",
	PAGES=	"074017",
	YEAR=	"2015",
}

@ARTICLE{CMS:2008xjf,
	AUTHOR=	"{CMS Collaboration}",
	TITLE=	"The {CMS} experiment at the {CERN} {LHC}",
	DOI=	"10.1088/1748-0221/3/08/S08004",
	JOURNAL=	"JINST",
	VOLUME=	"3",
	PAGES=	"S08004",
	YEAR=	"2008",
}

@ARTICLE{CMS:2023gfb,
	AUTHOR=	"{CMS Collaboration}",
	TITLE=	"Development of the {CMS} detector for the {CERN LHC Run 3}",
	DOI=	"10.1088/1748-0221/19/05/P05064",
	JOURNAL=	"JINST",
	VOLUME=	"19",
	PAGES=	"P05064",
	YEAR=	"2024",
}

@ARTICLE{CMS:2020cmk,
	AUTHOR=	"{CMS Collaboration}",
	TITLE=	"{Performance of the CMS Level-1 trigger in proton-proton collisions at $\sqrt{s} = 13$\,TeV}",
	JOURNAL=	"JINST",
	VOLUME=	"15",
	PAGES=	"P10017",
	YEAR=	"2020",
	DOI=	"10.1088/1748-0221/15/10/P10017",
	EPRINT=	"2006.10165",
	ARCHIVEPREFIX=	"arXiv",
	PRIMARYCLASS=	"hep-ex",
	REPORTNUMBER=	"CMS-TRG-17-001, CERN-EP-2020-065",
}

@ARTICLE{CMS:2016ngn,
	AUTHOR=	"{CMS Collaboration}",
	TITLE=	"{The CMS} trigger system",
	JOURNAL=	"JINST",
	VOLUME=	"12",
	PAGES=	"P01020",
	DOI=	"10.1088/1748-0221/12/01/P01020",
	YEAR=	"2017",
	EPRINT=	"1609.02366",
	ARCHIVEPREFIX=	"arXiv",
	PRIMARYCLASS=	"physics.ins-det",
	REPORTNUMBER=	"CMS-TRG-12-001, CERN-EP-2016-160",
	SLACCITATION=	"%%CITATION = ARXIV:1609.02366;%%",
}

@ARTICLE{CMS:2024aqx,
	AUTHOR=	"{CMS Collaboration}",
	TITLE=	"Performance of the {CMS} high-level trigger during {LHC} {Run} 2",
	EPRINT=	"2410.17038",
	ARCHIVEPREFIX=	"arXiv",
	PRIMARYCLASS=	"physics.ins-det",
	REPORTNUMBER=	"CMS-TRG-19-001, CERN-EP-2024-259",
	DOI=	"10.1088/1748-0221/19/11/P11021",
	JOURNAL=	"JINST",
	VOLUME=	"19",
	PAGES=	"P11021",
	YEAR=	"2024",
}

@ARTICLE{CMS:2020uim,
	AUTHOR=	"{CMS Collaboration}",
	TITLE=	"Electron and photon reconstruction and identification with the {CMS} experiment at the {CERN} {LHC}",
	EPRINT=	"2012.06888",
	JOURNAL=	"JINST",
	VOLUME=	"16",
	PAGES=	"P05014",
	YEAR=	"2021",
	ARCHIVEPREFIX=	"arXiv",
	PRIMARYCLASS=	"hep-ex",
	REPORTNUMBER=	"CMS-EGM-17-001, CERN-EP-2020-219",
	DOI=	"10.1088/1748-0221/16/05/P05014",
}

@ARTICLE{CMS:2018rym,
	AUTHOR=	"{CMS Collaboration}",
	TITLE=	"{Performance of the CMS muon detector and muon reconstruction with proton-proton collisions at $\sqrt{s}=$ 13 TeV}",
	EPRINT=	"1804.04528",
	ARCHIVEPREFIX=	"arXiv",
	PRIMARYCLASS=	"physics.ins-det",
	REPORTNUMBER=	"CMS-MUO-16-001, CERN-EP-2018-058",
	DOI=	"10.1088/1748-0221/13/06/P06015",
	JOURNAL=	"JINST",
	VOLUME=	"13",
	PAGES=	"P06015",
	YEAR=	"2018",
}

@ARTICLE{CMS:2014pgm,
	AUTHOR=	"{CMS Collaboration}",
	TITLE=	"{Description} and performance of track and primary-vertex reconstruction with the {CMS} tracker",
	EPRINT=	"1405.6569",
	ARCHIVEPREFIX=	"arXiv",
	PRIMARYCLASS=	"physics.ins-det",
	REPORTNUMBER=	"CMS-TRK-11-001, CERN-PH-EP-2014-070",
	DOI=	"10.1088/1748-0221/9/10/P10009",
	JOURNAL=	"JINST",
	VOLUME=	"9",
	PAGES=	"P10009",
	YEAR=	"2014",
}

@ARTICLE{CMS:2017yfk,
	AUTHOR=	"{CMS Collaboration}",
	TITLE=	"Particle-flow reconstruction and global event description with the {CMS} detector",
	JOURNAL=	"JINST",
	VOLUME=	"12",
	YEAR=	"2017",
	PAGES=	"P10003",
	DOI=	"10.1088/1748-0221/12/10/P10003",
	EPRINT=	"1706.04965",
	ARCHIVEPREFIX=	"arXiv",
	PRIMARYCLASS=	"physics.ins-det",
	REPORTNUMBER=	"CMS-PRF-14-001, CERN-EP-2017-110",
	SLACCITATION=	"%%CITATION = ARXIV:1706.04965;%%",
}

@ARTICLE{CMS:2019ctu,
	AUTHOR=	"{CMS Collaboration}",
	TITLE=	"Performance of missing transverse momentum reconstruction in proton-proton collisions at $\sqrt{s} = 13$\,{TeV} using the {CMS} detector",
	JOURNAL=	"JINST",
	VOLUME=	"14",
	YEAR=	"2019",
	PAGES=	"P07004",
	DOI=	"10.1088/1748-0221/14/07/P07004",
	EPRINT=	"1903.06078",
	ARCHIVEPREFIX=	"arXiv",
	PRIMARYCLASS=	"hep-ex",
	REPORTNUMBER=	"CMS-JME-17-001, CERN-EP-2018-335",
	SLACCITATION=	"%%CITATION = ARXIV:1903.06078;%%",
}

@ARTICLE{MadGraph5_aMCatNLO,
	AUTHOR=	"Alwall, J. and Frederix, R. and Frixione, S. and Hirschi, V. and Maltoni, F. and Mattelaer, O. and Shao, H. S. and Stelzer, T. and Torrielli, P. and Zaro, M.",
	TITLE=	"The automated computation of tree-level and next-to-leading order differential cross sections, and their matching to parton shower simulations",
	JOURNAL=	"JHEP",
	VOLUME=	"07",
	YEAR=	"2014",
	PAGES=	"079",
	DOI=	"10.1007/JHEP07(2014)079",
	EPRINT=	"1405.0301",
	ARCHIVEPREFIX=	"arXiv",
	PRIMARYCLASS=	"hep-ph",
}

@ARTICLE{Frederix:2012ps,
	AUTHOR=	"Frederix, Rikkert and Frixione, Stefano",
	TITLE=	"{Merging} meets matching in {MC@NLO}",
	JOURNAL=	"JHEP",
	VOLUME=	"12",
	YEAR=	"2012",
	PAGES=	"061",
	DOI=	"10.1007/JHEP12(2012)061",
	EPRINT=	"1209.6215",
	ARCHIVEPREFIX=	"arXiv",
	PRIMARYCLASS=	"hep-ph",
	REPORTNUMBER=	"CERN-PH-TH-2012-247, ZU-TH-21-12",
	SLACCITATION=	"%%CITATION = ARXIV:1209.6215;%%",
}

@ARTICLE{Alwall:2007fs,
	AUTHOR=	"Alwall, Johan and others",
	TITLE=	"{Comparative study of various algorithms for the merging of parton showers and matrix elements in hadronic collisions}",
	JOURNAL=	"Eur. Phys. J. C",
	VOLUME=	"53",
	YEAR=	"2008",
	PAGES=	"473",
	DOI=	"10.1140/epjc/s10052-007-0490-5",
	EPRINT=	"0706.2569",
	ARCHIVEPREFIX=	"arXiv",
	PRIMARYCLASS=	"hep-ph",
	REPORTNUMBER=	"SLAC-PUB-12604, CERN-PH-TH-2007-066, LU-TP-07-13, KA-TP-06-2007, DCPT-07-62, IPPP-07-31",
	SLACCITATION=	"%%CITATION = ARXIV:0706.2569;%%",
}

@ARTICLE{CMS:2020lrr,
	AUTHOR=	"{CMS Collaboration}",
	TITLE=	"{Search for new physics in top quark production with additional leptons in proton-proton collisions at $\sqrt{s} = $ 13 TeV using effective field theory}",
	EPRINT=	"2012.04120",
	ARCHIVEPREFIX=	"arXiv",
	PRIMARYCLASS=	"hep-ex",
	REPORTNUMBER=	"CMS-TOP-19-001, CERN-EP-2020-211",
	DOI=	"10.1007/JHEP03(2021)095",
	JOURNAL=	"JHEP",
	VOLUME=	"03",
	PAGES=	"095",
	YEAR=	"2021",
}

@ARTICLE{Belvedere:2024nzh,
	AUTHOR=	"Belvedere, Alberto and others",
	TITLE=	"{LHC EFT WG} note: {SMEFT} predictions, event reweighting, and simulation",
	EPRINT=	"2406.14620",
	ARCHIVEPREFIX=	"arXiv",
	PRIMARYCLASS=	"hep-ph",
	REPORTNUMBER=	"CERN-LHCEFTWG-2024-001, CERN-LPCC-2024-001, CERN-LPCC-2024-001",
	DOI=	"10.21468/SciPostPhysCommRep.4",
	PAGES=	"1",
	YEAR=	"2024",
	VOLUME=	"4",
	JOURNAL=	"SciPost Phys. Comm. Repts.",
}

@ARTICLE{CMS:2024dzv,
	AUTHOR=	"{CMS Collaboration}",
	TITLE=	"{Search for baryon number violation in top quark production and decay using proton-proton collisions at $\sqrt{s}$ = 13 TeV}",
	EPRINT=	"2402.18461",
	ARCHIVEPREFIX=	"arXiv",
	PRIMARYCLASS=	"hep-ex",
	REPORTNUMBER=	"CMS-TOP-22-003, CERN-EP-2024-027",
	DOI=	"10.1103/PhysRevLett.132.241802",
	JOURNAL=	"Phys. Rev. Lett.",
	VOLUME=	"132",
	PAGES=	"241802",
	YEAR=	"2024",
}

@ARTICLE{POWHEG1,
	AUTHOR=	"Nason, Paolo",
	TITLE=	"A new method for combining {NLO} {QCD} with shower {Monte Carlo} algorithms",
	JOURNAL=	"JHEP",
	VOLUME=	"11",
	YEAR=	"2004",
	PAGES=	"040",
	DOI=	"10.1088/1126-6708/2004/11/040",
	EPRINT=	"hep-ph/0409146",
	ARCHIVEPREFIX=	"arXiv",
	PRIMARYCLASS=	"hep-ph",
}

@ARTICLE{POWHEG2,
	AUTHOR=	"Frixione, Stefano and Nason, Paolo and Oleari, Carlo",
	TITLE=	"Matching {NLO} {QCD} computations with parton shower simulations: the {\POWHEG} method",
	JOURNAL=	"JHEP",
	VOLUME=	"11",
	YEAR=	"2007",
	PAGES=	"070",
	DOI=	"10.1088/1126-6708/2007/11/070",
	EPRINT=	"0709.2092",
	ARCHIVEPREFIX=	"arXiv",
	PRIMARYCLASS=	"hep-ph",
}

@ARTICLE{POWHEG3,
	AUTHOR=	"Alioli, Simone and Nason, Paolo and Oleari, Carlo and Re, Emanuele",
	TITLE=	"A general framework for implementing {NLO} calculations in shower {Monte Carlo} programs: the {\POWHEG} \textsc{box}",
	JOURNAL=	"JHEP",
	VOLUME=	"06",
	YEAR=	"2010",
	PAGES=	"043",
	DOI=	"10.1007/JHEP06(2010)043",
	EPRINT=	"1002.2581",
	ARCHIVEPREFIX=	"arXiv",
	PRIMARYCLASS=	"hep-ph",
}

@ARTICLE{POWHEG_VV,
	AUTHOR=	"Melia, Tom and Nason, Paolo and Rontsch, Raoul and Zanderighi, Giulia",
	TITLE=	"${\PWp\PWm}$, ${\PW\PZ}$ and ${\PZ\PZ}$ production in the {\POWHEG} \textsc{box}",
	JOURNAL=	"JHEP",
	VOLUME=	"11",
	YEAR=	"2011",
	PAGES=	"078",
	DOI=	"10.1007/JHEP11(2011)078",
	EPRINT=	"1107.5051",
	ARCHIVEPREFIX=	"arXiv",
	PRIMARYCLASS=	"hep-ph",
}

@ARTICLE{Frixione:2007nw,
	AUTHOR=	"Frixione, S. and Ridolfi, G. and Nason, P.",
	TITLE=	"A positive-weight next-to-leading-order {Monte Carlo} for heavy flavour hadroproduction",
	EPRINT=	"0707.3088",
	ARCHIVEPREFIX=	"arXiv",
	PRIMARYCLASS=	"hep-ph",
	DOI=	"10.1088/1126-6708/2007/09/126",
	JOURNAL=	"JHEP",
	VOLUME=	"09",
	PAGES=	"126",
	YEAR=	"2007",
}

@ARTICLE{Sjostrand:2014zea,
	AUTHOR=	"Sj{\"o}strand, Torbj{\"o}rn and Ask, Stefan and Christiansen, Jesper R. and Corke, Richard and Desai, Nishita and Ilten, Philip and Mrenna, Stephen and Prestel, Stefan and Rasmussen, Christine O. and Skands, Peter Z.",
	TITLE=	"An introduction to {PYTHIA} 8.2",
	JOURNAL=	"Comput. Phys. Commun.",
	VOLUME=	"191",
	YEAR=	"2015",
	PAGES=	"159",
	DOI=	"10.1016/j.cpc.2015.01.024",
	EPRINT=	"1410.3012",
	ARCHIVEPREFIX=	"arXiv",
	PRIMARYCLASS=	"hep-ph",
	REPORTNUMBER=	"LU-TP-14-36, MCNET-14-22, CERN-PH-TH-2014-190, FERMILAB-PUB-14-316-CD, DESY-14-178, SLAC-PUB-16122",
	SLACCITATION=	"%%CITATION = ARXIV:1410.3012;%%",
}

@ARTICLE{Sirunyan:2019dfx,
	AUTHOR=	"{CMS Collaboration}",
	TITLE=	"Extraction and validation of a new set of {CMS} {PYTHIA8} tunes from underlying-event measurements",
	EPRINT=	"1903.12179",
	ARCHIVEPREFIX=	"arXiv",
	PRIMARYCLASS=	"hep-ex",
	REPORTNUMBER=	"CMS-GEN-17-001, CERN-EP-2019-007",
	DOI=	"10.1140/epjc/s10052-019-7499-4",
	JOURNAL=	"Eur. Phys. J. C",
	VOLUME=	"80",
	PAGES=	"4",
	YEAR=	"2020",
}

@ARTICLE{Ball:2017nwa,
	AUTHOR=	"Ball, Richard D. and others",
	TITLE=	"Parton distributions from high-precision collider data",
	COLLABORATION=	"NNPDF",
	JOURNAL=	"Eur. Phys. J. C",
	VOLUME=	"77",
	YEAR=	"2017",
	PAGES=	"663",
	DOI=	"10.1140/epjc/s10052-017-5199-5",
	EPRINT=	"1706.00428",
	ARCHIVEPREFIX=	"arXiv",
	PRIMARYCLASS=	"hep-ph",
	REPORTNUMBER=	"CAVENDISH-HEP-17-06, CERN-TH-2017-077, EDINBURGH-2017-08, NIKHEF-2017-006, OUTP-17-04P, TIF-UNIMI-2017-3",
	SLACCITATION=	"%%CITATION = ARXIV:1706.00428;%%",
}

@ARTICLE{Agostinelli:2002hh,
	AUTHOR=	"Agostinelli, S. and others",
	COLLABORATION=	"GEANT4",
	TITLE=	"{\GEANTfour}---a simulation toolkit",
	JOURNAL=	"Nucl. Instrum. Meth. A",
	VOLUME=	"506",
	YEAR=	"2003",
	PAGES=	"250",
	DOI=	"10.1016/S0168-9002(03)01368-8",
	SLACCITATION=	"%%CITATION = NUIMA,A506,250;%%",
}

@ARTICLE{Bols:2020bkb,
	AUTHOR=	"Bols, Emil and Kieseler, Jan and Verzetti, Mauro and Stoye, Markus and Stakia, Anna",
	TITLE=	"Jet flavour classification using {DeepJet}",
	EPRINT=	"2008.10519",
	ARCHIVEPREFIX=	"arXiv",
	PRIMARYCLASS=	"hep-ex",
	DOI=	"10.1088/1748-0221/15/12/P12012",
	JOURNAL=	"JINST",
	VOLUME=	"15",
	PAGES=	"P12012",
	YEAR=	"2020",
}

@TECHREPORT{CMS-DP-2023-005,
	AUTHOR=	"{CMS Collaboration}",
	TITLE=	"Performance summary of {AK4} jet {\PQb} tagging with data from proton-proton collisions at {13\TeV} with the {CMS} detector",
	YEAR=	"2023",
	URL=	"https://cds.cern.ch/record/2854609",
	TYPE=	"CMS Detector Performance Note",
	NUMBER=	"CMS-DP-2023-005",
}

@ARTICLE{Cacciari:2011ma,
	AUTHOR=	"Cacciari, Matteo and Salam, Gavin P. and Soyez, Gregory",
	TITLE=	"{FastJet} user manual",
	JOURNAL=	"Eur. Phys. J. C",
	VOLUME=	"72",
	PAGES=	"1896",
	DOI=	"10.1140/epjc/s10052-012-1896-2",
	YEAR=	"2012",
	EPRINT=	"1111.6097",
	ARCHIVEPREFIX=	"arXiv",
	PRIMARYCLASS=	"hep-ph",
	REPORTNUMBER=	"CERN-PH-TH-2011-297",
	SLACCITATION=	"%%CITATION = ARXIV:1111.6097;%%",
}

@ARTICLE{Hoecker:2007ht,
	AUTHOR=	"Hoecker, Andreas and Speckmayer, Peter and Stelzer, Joerg and Therhaag, Jan and von Toerne, Eckhard and Voss, Helge",
	TITLE=	"{TMVA}: {Toolkit} for {Multivariate Data Analysis}",
	JOURNAL=	"PoS",
	VOLUME=	"ACAT",
	YEAR=	"2007",
	PAGES=	"040",
	EPRINT=	"physics/0703039",
	ARCHIVEPREFIX=	"arXiv",
}

@ARTICLE{Jackson:2017gcy,
	AUTHOR=	"Jackson, Paul and Rogan, Christopher",
	TITLE=	"{Recursive Jigsaw Reconstruction}: {HEP} event analysis in the presence of kinematic and combinatoric ambiguities",
	EPRINT=	"1705.10733",
	ARCHIVEPREFIX=	"arXiv",
	PRIMARYCLASS=	"hep-ph",
	DOI=	"10.1103/PhysRevD.96.112007",
	JOURNAL=	"Phys. Rev. D",
	VOLUME=	"96",
	PAGES=	"112007",
	YEAR=	"2017",
}

@ARTICLE{CMS:2020mpn,
	AUTHOR=	"{CMS Collaboration}",
	TITLE=	"Measurement of the {Higgs} boson production rate in association with top quarks in final states with electrons, muons, and hadronically decaying tau leptons at $\sqrt{s}={13\TeV}$",
	EPRINT=	"2011.03652",
	ARCHIVEPREFIX=	"arXiv",
	PRIMARYCLASS=	"hep-ex",
	DOI=	"10.1140/epjc/s10052-021-09014-x",
	JOURNAL=	"Eur. Phys. J. C",
	VOLUME=	"81",
	PAGES=	"378",
	YEAR=	"2021",
}

@ARTICLE{CMS:2017moi,
	AUTHOR=	"{CMS Collaboration}",
	TITLE=	"Search for electroweak production of charginos and neutralinos in multilepton final states in proton-proton collisions at $\sqrt{s}={13\TeV}$",
	EPRINT=	"1709.05406",
	ARCHIVEPREFIX=	"arXiv",
	PRIMARYCLASS=	"hep-ex",
	DOI=	"10.1007/JHEP03(2018)166",
	JOURNAL=	"JHEP",
	VOLUME=	"03",
	PAGES=	"166",
	YEAR=	"2018",
}

@ARTICLE{CMS:2022tkv,
	AUTHOR=	"{CMS Collaboration}",
	TITLE=	"{Measurement of the cross section of top quark-antiquark pair production in association with a W boson in proton-proton collisions at $ \sqrt{s} $ = 13 TeV}",
	EPRINT=	"2208.06485",
	ARCHIVEPREFIX=	"arXiv",
	PRIMARYCLASS=	"hep-ex",
	REPORTNUMBER=	"CMS-TOP-21-011, CERN-EP-2022-168",
	DOI=	"10.1007/JHEP07(2023)219",
	JOURNAL=	"JHEP",
	VOLUME=	"07",
	PAGES=	"219",
	YEAR=	"2023",
}

@ARTICLE{ATLAS:2024moy,
	AUTHOR=	"{ATLAS Collaboration}",
	TITLE=	"{Measurement of the total and differential cross-sections of $ t\overline{t}W $ production in pp collisions at $ \sqrt{s} $ = 13 TeV with the ATLAS detector}",
	EPRINT=	"2401.05299",
	ARCHIVEPREFIX=	"arXiv",
	PRIMARYCLASS=	"hep-ex",
	REPORTNUMBER=	"CERN-EP-2023-260",
	DOI=	"10.1007/JHEP05(2024)131",
	JOURNAL=	"JHEP",
	VOLUME=	"05",
	PAGES=	"131",
	YEAR=	"2024",
	NOTE=	"[Erratum: \href{https://doi.org/10.1007/JHEP11(2025)127}{JHEP 11 (2025) 127}]",
}

@ARTICLE{CMS:2024onh,
	AUTHOR=	"{CMS Collaboration}",
	TITLE=	"The {CMS} statistical analysis and combination tool: \textsc{Combine}",
	EPRINT=	"2404.06614",
	ARCHIVEPREFIX=	"arXiv",
	PRIMARYCLASS=	"physics.data-an",
	DOI=	"10.1007/s41781-024-00121-4",
	JOURNAL=	"Comput. Softw. Big Sci.",
	VOLUME=	"8",
	PAGES=	"19",
	YEAR=	"2024",
}

@MISC{hepdata,
	HOWPUBLISHED=	"{HEPD}ata record for this analysis",
	DOI=	"10.17182/hepdata.172985",
	YEAR=	"2026",
}

@ARTICLE{Wilks:1938dza,
	AUTHOR=	"Wilks, S. S.",
	TITLE=	"{The} large-sample distribution of the likelihood ratio for testing composite hypotheses",
	JOURNAL=	"Ann. Math. Stat.",
	VOLUME=	"9",
	PAGES=	"60",
	YEAR=	"1938",
	DOI=	"10.1214/aoms/1177732360",
}

@ARTICLE{Bernlochner:2022oiw,
	AUTHOR=	"Bernlochner, Florian Urs and Fry, Daniel C. and Menary, Stephen Burns and Persson, Eric",
	TITLE=	"{Cover} your bases: asymptotic distributions of the profile likelihood ratio when constraining effective field theories in high-energy physics",
	EPRINT=	"2207.01350",
	ARCHIVEPREFIX=	"arXiv",
	PRIMARYCLASS=	"physics.data-an",
	DOI=	"10.21468/SciPostPhysCore.6.1.013",
	JOURNAL=	"SciPost Phys. Core",
	VOLUME=	"6",
	PAGES=	"013",
	YEAR=	"2023",
}
\cleardoublepage \appendix\section{The CMS Collaboration \label{app:collab}}\begin{sloppypar}\hyphenpenalty=5000\widowpenalty=500\clubpenalty=5000\cmsinstitute{Yerevan Physics Institute, Yerevan, Armenia}
{\tolerance=6000
A.~Gevorgyan\cmsorcid{0000-0003-2751-9489}, A.~Hayrapetyan, V.~Makarenko\cmsorcid{0000-0002-8406-8605}, A.~Tumasyan\cmsAuthorMark{1}\cmsorcid{0009-0000-0684-6742}
\par}
\cmsinstitute{Institut f\"{u}r Hochenergiephysik, Vienna, Austria}
{\tolerance=6000
P.S.~Hussain\cmsorcid{0000-0002-4825-5278}, M.~Sonawane\cmsorcid{0000-0003-0510-7010}
\par}
\cmsinstitute{Marietta Blau Institute for Particle Physics, Vienna, Austria}
{\tolerance=6000
W.~Adam\cmsorcid{0000-0001-9099-4341}, L.~Benato\cmsorcid{0000-0001-5135-7489}, T.~Bergauer\cmsorcid{0000-0002-5786-0293}, M.~Dragicevic\cmsorcid{0000-0003-1967-6783}, S.~Gundacker\cmsorcid{0000-0003-2087-3266}, A.K.~Guven\cmsorcid{0009-0004-5670-5138}, M.~Jeitler\cmsAuthorMark{2}\cmsorcid{0000-0002-5141-9560}, N.~Krammer\cmsorcid{0000-0002-0548-0985}, A.~Li\cmsorcid{0000-0002-4547-116X}, D.~Liko\cmsorcid{0000-0002-3380-473X}, M.~Matthewman, A.~Pfeiffer\cmsorcid{0000-0001-5328-448X}, J.~Schieck\cmsAuthorMark{2}\cmsorcid{0000-0002-1058-8093}, R.~Sch\"{o}fbeck\cmsAuthorMark{2}\cmsorcid{0000-0002-2332-8784}, M.~Shooshtari\cmsorcid{0009-0004-8882-4887}, N.~Van~Den~Bossche\cmsorcid{0000-0003-2973-4991}, W.~Waltenberger\cmsorcid{0000-0002-6215-7228}, C.E.~Wulz\cmsAuthorMark{2}\cmsorcid{0000-0001-9226-5812}
% ...
\par}
$^{1}$Also at Yerevan State University, Yerevan, Armenia\\
$^{2}$Also at Technische Universit\"{a}t Wien, Vienna, Austria\\
% ...
\end{sloppypar}
\end{document}